%% file: sample-sigconf.tex
\documentclass[sigplan,9pt]{acmart}

\usepackage{booktabs} 

\acmYear{2019}\copyrightyear{2019}
\setcopyright{licensedothergov}
\acmConference[EdgeSys '19]{EdgeSys '19: International Workshop on Edge Systems, Analytics and Networking}{March 25, 2019}{Dresden, Germany}
\acmBooktitle{EdgeSys '19: International Workshop on Edge Systems, Analytics and Networking, March 25, 2019, Dresden, Germany}

\acmDOI{10.1145/3301418.3313947}

\acmISBN{978-1-4503-6275-7/19/03}

\acmPrice{15.00}

\begin{document}
\title{Checkpointing and Migration of IoT Edge Functions}

\author{Pekka Karhula}
\authornote{The work was performed during a research visit at Columbia University}
\affiliation{%
  \institution{VTT Technical Research Centre of Finland Ltd}
  \streetaddress{Kaitov\"ayl\"a}
  \city{Oulu}
  \country{Finland}
  \postcode{90570}
}

\email{pekka.karhula@vtt.fi}

\author{Jan Janak}
\affiliation{%
  \institution{Columbia University}
  \city{New York City}
  \state{NY}
  \postcode{10027}
}
\email{janakj@cs.columbia.edu}

\author{Henning Schulzrinne}
\affiliation{%
  \institution{Columbia University}
  \city{New York City}
  \state{NY}
  \postcode{10027}
}
\email{hgs@cs.columbia.edu}



\begin{abstract}
The serverless and functions as a service (FaaS) paradigms are currently trending among cloud providers and are now increasingly being applied to the network edge, and to the Internet of Things (IoT) devices. The benefits include reduced latency for communication, less network traffic and increased privacy for data processing. However, there are challenges as IoT devices have limited resources for running multiple simultaneous containerized functions, and also FaaS does not typically support long-running functions. Our implementation utilizes Docker and CRIU for checkpointing and suspending long-running blocking functions. The results show that checkpointing is slightly slower than regular Docker pause, but it saves memory and allows for more long-running functions to be run on an IoT device. Furthermore, the resulting checkpoint files are small, hence they are suitable for live migration and backing up stateful functions, therefore improving availability and reliability of the system. 
\end{abstract}

%
%



\maketitle

\input{content}

\bibliographystyle{ACM-Reference-Format}
\bibliography{refs}

\end{document}

%% file: content.tex
\section{Introduction}

The serverless paradigm is currently an emerging trend among cloud service providers. The key idea is that clients do not have to worry about the provisioning, maintenance or scaling of their service as all that is done by the serverless providers. Instead, more focus and time can be put into developing code that is run in the serverless platform. Another big trend in the serverless paradigm is the Function as a Service (FaaS) execution model. It further refines the microservice structure into functions that are triggered by events such as HTTP requests or video streams. 


In addition to the cloud, the FaaS execution model is attractive for event-driven tasks at the network edge. 
The benefits include lower latency, privacy, reduced network traffic and energy efficiency. IoT devices that can run Linux containers, could naturally be used to provide FaaS within an IoT network. IoT applications are likely to benefit from a more flexible execution model that supports long-running (stateful) functions.

The FaaS model has one crucial drawback: it does not typically support long computations. Furthermore, the functions are stateless, meaning that the computation cannot be paused and continued later. Directly applying this approach to IoT devices would be meaningless, as we do not want to abandon a long computation and waste resources due to a time limit. However, running multiple long-running function instances may be difficult on memory-constrained IoT due to the memory requirements of Docker. A novel way of managing inactive function containers is needed to make long-running functions feasible on IoT devices.

To tackle this issue, we use Checkpoint/Restore In Userspace (CRIU) \cite{criu} together with Docker for preserving the function states through checkpointing. With this method, we were able to checkpoint and restore TCP/IP sockets and sleeping functions, which is essential for enabling long-running functions. Also, our tests show that this method can be used for live migrating container states from one device to another. Possible use cases include fault-tolerant IoT systems and resource offloading. 


The rest of the paper is organized as follows. In Section \ref{background}, background and related work are presented. Section \ref{lambda} discusses FaaS and its applicability to IoT and edge computing. Section \ref{results} provides an overview of the early implementation and performance results. The paper is concluded in Section \ref{conclusion}.

\section{Background}
\label{background}

The idea behind FaaS is fairly straightforward: let the user write and run an event-driven program without the need to provision infrastructure for it. As of January 2019, the major FaaS platforms include Amazon AWS Lambda~\cite{amazonlambda}, Google Cloud Functions~\cite{googlecloud} and Microsoft Azure Functions~\cite{microsoftazure}. In addition, Apache OpenWhisk \cite{openwhisk} and OpenLambda \cite{hendrickson2016serverless} are two well-known open source serverless implementations.

All major FaaS platforms support program activation by HTTP requests. Upon receiving an HTTP request, the framework allocates resources for the program and runs the program, giving it access to the HTTP request that triggered it. After the program has sent a response, the process is terminated. Most frameworks support additional triggers from cloud-specific subsystems (e.g., a database or publish/subscribe framework). Typically, the program is developed in a high-level programming language (JavaScript or Python) on top of a framework provided by the cloud service provider.

When the same FaaS program is invoked multiple times, there is no guarantee that it will be invoked on the same machine. For that reason, the typical FaaS program is stateless. If the program needs to keep state from one invocation to the next, the state needs to be either supplied by the client, or stored in an external database.

Before a FaaS program can be invoked on a machine for the first time, its container must be created and started from the beginning. This is called a cold start. Warm start, on the other hand, has the container already loaded in the memory and ready to serve requests. While cold starts make auto-scaling possible, at the same time they make stateful FaaS programs difficult to implement. In terms of response times, a warm start is preferable over a cold start. However, maintaining large numbers of inactive pre-loaded containers for warm starts tends to be resource-intensive.

When deploying a FaaS program on a device at the network edge, the need to conserve memory arises quickly due to the constrained nature of edge devices. Thus, deploying FaaS programs at the network edge requires a careful trade-off between efficiency and resource usage. Process pausing and checkpointing are two promising techniques that may help conserve memory on constrained edge devices running FaaS containers. Rather than cold starting a FaaS program on each invocation, the program's memory could be saved to disk when memory on the device runs low and re-loaded on the next invocation. Apart from conserving memory, the checkpointing technique also enables stateful FaaS programs useful in IoT applications.


\subsection{Related Work}
\label{related}

 In \cite{wang2018peeking}, Wang et al. conducted a measurement study of Amazon AWS Lambda, Microsoft Azure and Google Cloud Functions. Their study provided more insight on how these three major FaaS providers handle resource scheduling, utilization and isolation. The popular FaaS and Platform-as-a-Service (PaaS) models have also been applied to the IoT edge network. Mehta et al. \cite{Mehta2017} implemented an IoT framework based on a serverless architecture and FaaS, which they refer to as actor-as-a-service. The actors work much the same way as functions in FaaS, but in the IoT scenario they may also represent sensors and actuators. 
 There are some limitations in FaaS as noted in \cite{hellerstein2018serverless, wang2018peeking}. The functions are non-addressable and short lived, which make it impossible to call a specific function instance or guarantee that repeated requests are delivered to the same function instance. 

Docker container technology performance in IoT devices has been evaluated in several studies \cite{Morabito2017, Kakakhel2018, bellavista:rpifeasibility}. The results show that running containers on Raspberry Pis is feasible in terms of resource usage and speed. Bellavista et al. \cite{bellavista:rpifeasibility} conducted a feasibility study of using Raspberry Pis and Docker in a fog computing scenario. In their experiment, they ran up to thirty concurrent containers and found out that the execution time is linearly dependent on the number of containers.


To deal with the memory contention, one could suspend the containers that are blocked waiting for network I/O. One way to implement this is using docker pause \cite{dockerpause} and the cgroup freezer \cite{cgroupfreezer}. However, results in \cite{hendrickson2016serverless} showed no difference in memory usage between a paused and a running container, which suggests that other approaches would be needed in order to save system memory.

Checkpointing could be a solution to the memory contention, as it will save the process state to disk. Chen \cite{Chen2015} conducted an evaluation of checkpointing with CRIU utilizing Docker-based micro-services on a desktop computer. The evaluation showed that checkpoint and restore times scaled linearly with the application image size. 
In \cite{Kakakhel2018}, live migration using CRIU and Docker was examined. Their result showed a 2.1x increase in processing time using live migration compared to algorithm execution without migration. They also experienced some transfer errors during the live migration. However, they transferred whole containers during the live migration, which naturally produces a lot of overhead especially on slower connections. Our idea is to maintain a set of base images in each of the IoT devices, so that most of the time we would need to transfer only the checkpoint files between the devices. Container migration has also been explored in \cite{Qiu2017}, where CRIU and Linux Containers (LXC) were combined, and in \cite{Mirhoseini2016}, which presented a framework for checkpointing long computations on low power devices. In \cite{Aissaoui2016}, Distributed MultiThreaded CheckPointing (DMTCP) was used to improve efficiency of Raspberry Pi based IoT devices in non-containerized applications.

The FaaS providers are also moving computation to the edge network. Examples include Amazon Lambda$@$Edge \cite{amazonlambdaedge} and Azure IoT Edge \cite{azureedge}. Azure IoT Edge shares the same underlying idea with our approach. When combined with Azure Durable functions extension \cite{durablefunctionsfw}, they also have support for stateful functions, where the framework manages state and restarts on behalf of the function. However, durable functions currently support only three programming languages whereas our approach is agnostic of the programming language or libraries. 

\section{Function as a Service on the IoT Edge}
\label{lambda}
In this section, we discuss the applicability of the FaaS model to programming IoT devices in the edge network. Consider an edge network with one or more IoT devices that can be programmed (out-of-band) to issue requests (HTTP or CoAP) upon the occurrence of some event. Let's assume we wish to invoke a function in response to the request.

An obvious first solution would be to setup an HTTP server in the edge network to start a container (e.g., Docker) with the function to handle the request. Once the function has terminated, the container is kept alive for a while and then it is terminated. The mapping from HTTP URLs to Docker containers would be stored in a configuration file. This kind of framework would be trivial to build using existing software.

The previous paragraph describes the most rudimentary FaaS framework implemented on a device in the edge network. Such a framework would run into memory limits of the device quickly. The maximum number of simultaneous containers the device could run is likely to be severely limited. The framework would need to employ aggressive container management and shutdown idle containers quickly. Also, functions, which run on shared infrastructure in the cloud, have time limits to minimize the cost. Such limits are unnecessary on an IoT device in the edge network that the user owns. Without the limits, the functions are more likely to be long-running. Long-running functions will exacerbate the simultaneous container problem.

Often, a function is long-running because it needs to wait for network I/O or a timer before it can terminate. Consider a function invoked by device A that waits for a notification from device B before it can send a response to device A. Such function might issue an HTTP request to device B and go to sleep until it receives a response from B. Running multiple long-running function containers on a constrained device may result in memory contention.

\subsection{Checkpointing}

Clearly, the FaaS model that works for cloud, would not work on constrained devices very well without some modifications. On constrained devices, we would want to minimize resource usage by avoiding cold starts, migrating computation to devices with low load, and pausing containers that are idle. One way to accomplish these objectives is to save the function and container state to disk, and continue execution when there is work to do, preferably on a device that is optimal for the task. 

The large ecosystem of Docker and performance results from prior research were factors that led us to choose the combination of Docker and CRIU for checkpointing in the IoT edge. We can use CRIU to save the state of the container to disk. The resulting checkpoint is basically a set of image files that can then be loaded by a local container or sent to a remote device.

\subsection{Motivating Scenarios}
\label{scenarios}

In addition to lowering resource usage, checkpointing could enable important IoT use cases that would be difficult to implement with traditional serverless and FaaS frameworks. The first example in this section is authorization, where checkpointing makes running containers more efficient by pausing inactive containers. The other two scenarios enabled by checkpointing are load balancing and fault-tolerant IoT, which are very important concepts in critical IoT systems.

\subsubsection{Authorization}
\label{auth}

\begin{figure}[h!]
    \centering
    \includegraphics[width=0.8\linewidth]{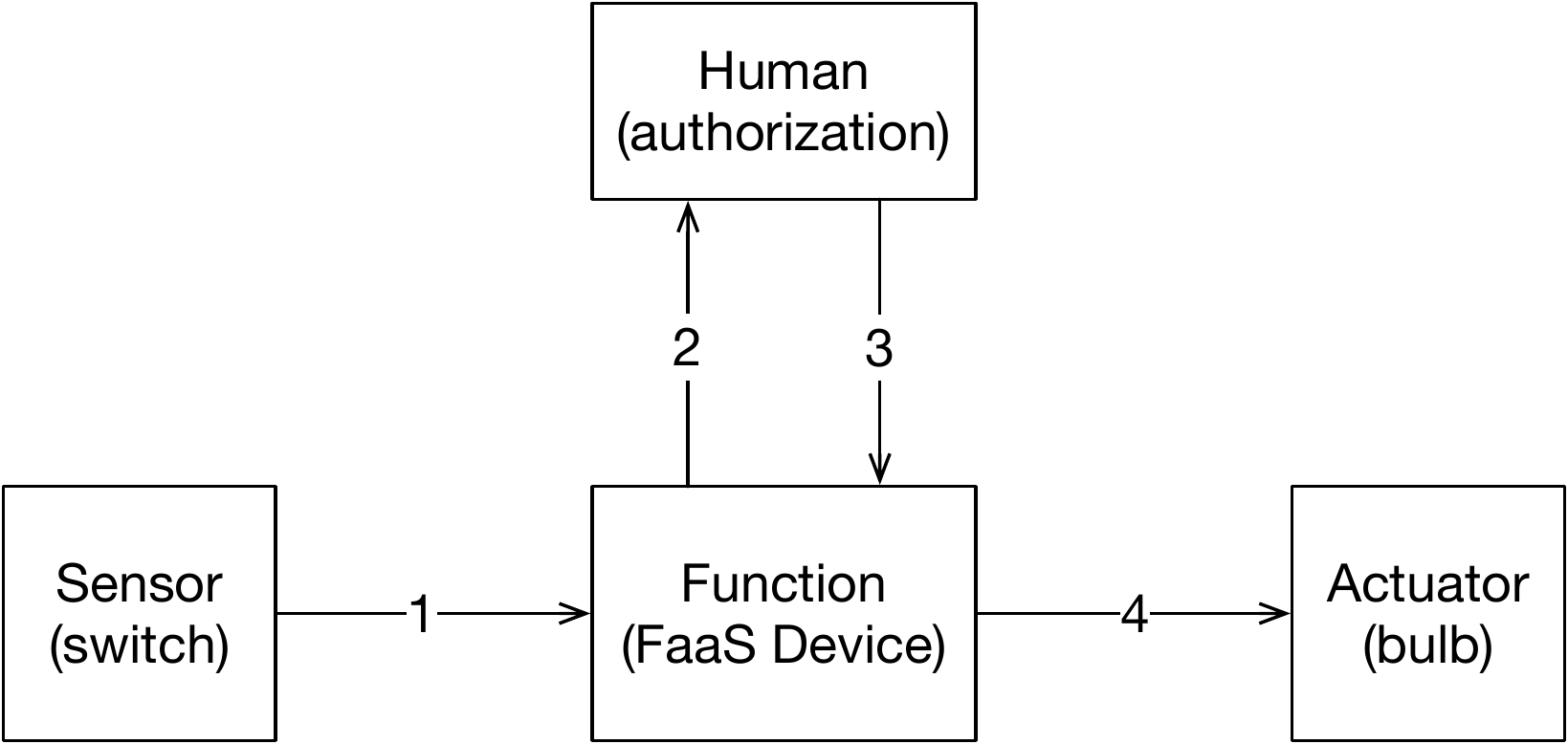}
    \caption{Example of a long-running function with blocking behavior}
    \label{fig:authorization}
\end{figure}

Consider the scenario illustrated in the Fig. \ref{fig:authorization}, where a networked light switch controls a networked light bulb. A third party must approve all requests to turn the light bulb off as a safety measure. The third party could be, e.g., an automated process that enforces a delay, or it could be manually operated by a human. In either case, it can take a long time to get a response back to the function.

The scenario could be implemented using a function deployed in the edge network. Instead of sending a request directly to the light bulb, the switch is re-configured to send the request to the function device. The function first obtains an authorization from the human. Once authorized, the function forwards the request to the light bulb.

Naturally, the scenario would be trivial to implement with a permanently installed and constantly running application. Such application would be, however, doing nothing and consuming memory for long periods of time. 
Also, this approach is not very scalable and maintainable.

A better approach is to use the FaaS execution model and activate the application upon receiving the request from the switch. This would work well if the function were stateless and idempotent. Unfortunately, that is not the case in our scenario. The framework could run the function until completion, however, since there is human involved, the time to completion may be long and hard to predict. What happens if the device runs out of memory? What if there are other short-term functions that need to be run while this one is waiting? Also, relying on the function staying idle for long periods of time is not good for reliability.

Yet another approach could be to extract the state out of the conversation between the function and the user, save it in a database, and periodically re-activate the function to poll the remote authorization server. This solution is, unfortunately, only possible if the protocol between the function and the remote server is well-known and exposed to the application. This is often not the case, e.g., when a third-party library is used.

Finally, the function could be written to use an external ``sleep proxy'' service (discussed later) to monitor its file descriptors while the function process is suspended. The sleep proxy would re-activate the function upon activity on any of the file descriptors.


\subsubsection{Load Balancing}
\label{loadbalance}

IoT devices may choose to offload tasks to other devices for better performance, lower energy usage, and reduced network traffic, among other metrics. If an IoT device is running near its full capacity, some of the tasks could be checkpointed and migrated into another device in the network. This naturally requires coordination among the IoT devices. The result would be improved availability and performance of the IoT system.

\subsubsection{Fault-tolerant IoT Systems}
\label{fault}

Checkpointing functionality could be used to backup the configuration of functions running on IoT devices periodically. Checkpointing is a common concept in traditional operating systems and applications in order to achieve fault-tolerant systems. However, it is not commonly used in IoT, which could also benefit from this concept. In IoT scenarios, the checkpoint images could be stored in the cloud or on an edge server and retrieved in case of device failure. The checkpointed backup function could also be continued on another device until it can be migrated back to the original device.

Fault-tolerance and reliability of IoT systems has received little attention from from the research community so far. The FaaS execution model together with process checkpointing might provide some of the building blocks for fault-tolerant IoT systems. 

\subsection{Sleep Proxy}

The checkpointing of functions brings some additional challenges. For example, how do we put containers to sleep and still know when to activate them? We would need to have some sort of external process that monitors file descriptors and incoming network connections. Before being checkpointed, the function submits a set of resource to be monitored to the external process and asks to be resumed when activity is detected on any of the resources. We call the external process a sleep proxy.

There is a number of ways to implement the sleep proxy service, depending on the architecture and characteristics of the function and its run-time environment. For example, NodeJS maintains a set of file descriptors to monitor for events using an external library called libuv. Thus, in NodeJS-based functions, one could extract the set of file descriptors to monitor from the library.

Regardless of the implementation details of the function being checkpointed, the sleep proxy pattern provides a general mechanism for the FaaS framework to learn about when to activate a particular function in case it needs to be checkpointed to alleviate memory contention. In addition, the sleep proxy pattern allows the function to modify the set of conditions under which it gets activated.


There needs to be only one sleep proxy process shared by all function containers. The implementation of sleep proxy will be complicated by the fact that file descriptors may need to be transferred across container boundaries. With this kind of mechanism, it would be possible to have a large number of long-running suspended functions "laying around" on a constrained device like the Raspberry Pi. Furthermore, the events that activate the function would not be limited to the statically configured HTTP requests, but the function could create its own activation conditions by waiting on network I/O.

\subsection{Limitations}

The FaaS execution model, as implemented by major cloud providers, has certain limitations. First of all, the activation condition, i.e., the set of rules that determine whether to invoke the function, must be static. Typically, activation conditions are configured by the user when creating a new function. The function itself cannot modify the activation condition, e.g., to install additional activation rules.

Before the program is run for the first time, the FaaS framework must provision a temporary execution environment for the program. That typically involves creating and initializing a container, setting up libraries, and so on. This process, referred to as "cold start" can take some time and may increase the latency of the first program invocation. The FaaS framework typically attempts to re-use the existing context for some time. It is unclear whether or not context reuse guarantees that the OS process running the function will be reused.

The FaaS program must be stateless and should be idempotent. All persistent data must be saved to an external database. This is needed to make it possible for the FaaS framework to run multiple instances of the program, e.g., for auto-scaling purposes. There are resource limits that constrain file system access, running time, and the memory available to the program. Furthermore, the program author has limited or no control over concurrency.

\section{Implementation and Preliminary Results}
\label{results}

The focus of this section is on the checkpointing functionality for FaaS programs. We use Raspberry Pi's for the experimental part as they represent well the computational capabilities of many IoT devices. These types of devices also have small form factor, which makes them useful in home and industrial gateways, as well as embeddable in vehicles and machinery.  

We have implemented a set of small NodeJS programs that are run inside containers in order to test the feasibility of the checkpoint functionality. In particular, we have set up a couple of tests that need to pass successfully in order to implement the scenarios that were introduced in Section \ref{scenarios}. We would want the framework to be able to enable the following functionality:

\begin{itemize}
    \item Checkpointing sleeping functions
    \item Checkpointing TCP/IP sockets
    \item Migrating to another device
\end{itemize}

In addition, we have run some performance related tests, as it is important for the feasibility of this approach that the tasks execute in reasonable time without using too many system resources.




\subsection{Setting up the Software}
\label{sec:setting-up-software}

CRIU functionality on Docker can be enabled through Docker's experimental features. Thus, the two are not fully integrated and the experimental status may cause some features to break in future releases. In fact, the checkpointing functionality has been faulty since Docker-ce 17.06 version.\footnote{https://github.com/moby/moby/issues/34601} There have been gradual fixes, but the current version (18.09) is still missing the functionality for dumping the checkpoint data to an arbitrary location in the filesystem. This functionality is very useful when one needs to migrate and continue the program in some other container or device. Docker-ce version 17.03.2 was selected for this implementation, as it is the last one where all the desired functionality exists.

The runtime environment was set up on a Raspberry Pi 2 Model B. At the time of experimentation, version 3.11 was the latest version of CRIU and it works well on the RPi as long as the Linux kernel prerequisites are met. The Linux kernel shipped with Raspbian did not have all the required kernel options enabled by default and so the kernel needed to be recompiled with the options listed in \cite{criukernel}. 

\subsection{Checkpointing a Long-running (sleeping) Function}

For this scenario, a simple JavaScript function simulating blocking behaviour of a long running function was created. The script was built into a Docker image containing the NodeJS dependencies. The steps for this experiment went as follows. 


\begin{enumerate}
    \item Run container
    \item Let it run for 30 seconds
    \item Checkpoint and suspend the container
    \item Wait another 30 seconds
    \item Continue from the checkpoint and check status from logs
\end{enumerate}

The container is run only half of the specified 60 seconds of the sleep time and then checkpointed and suspended. We want to know if this method can be used to exhaust the sleep timer without running the container in the background. After waiting another 30 seconds, the container is continued. The logs revealed that the sleep function had finished and the code continued running. This behavior is very useful, as our aim is to suspend functions that are sleeping or blocking for more optimal utilization of system resources. Naturally, utilizing this functionality in an IoT device would require some external program to keep track of the suspended containers, and to know when to wake them up. 

\subsection{TCP/IP Socket Checkpointing}

The previous section discussed checkpointing a function that is blocked by a timer with a predefined value. The blocking behavior can also be caused by a request that needs to be accepted by another IoT device or a person, like in the example in Section \ref{auth}. In that case, it might be useful to preserve also the TCP/IP connection when checkpointing.

In this case, an IoT device sends an HTTP request to a remote device operated by a human. A TCP/IP connection is created between the devices, but the response is delayed. We wanted to test the checkpointing of a container blocked in network I/O and resuming it again upon response arrival. 




\begin{figure}[h!]
    \centering
    \includegraphics[width=0.6\linewidth]{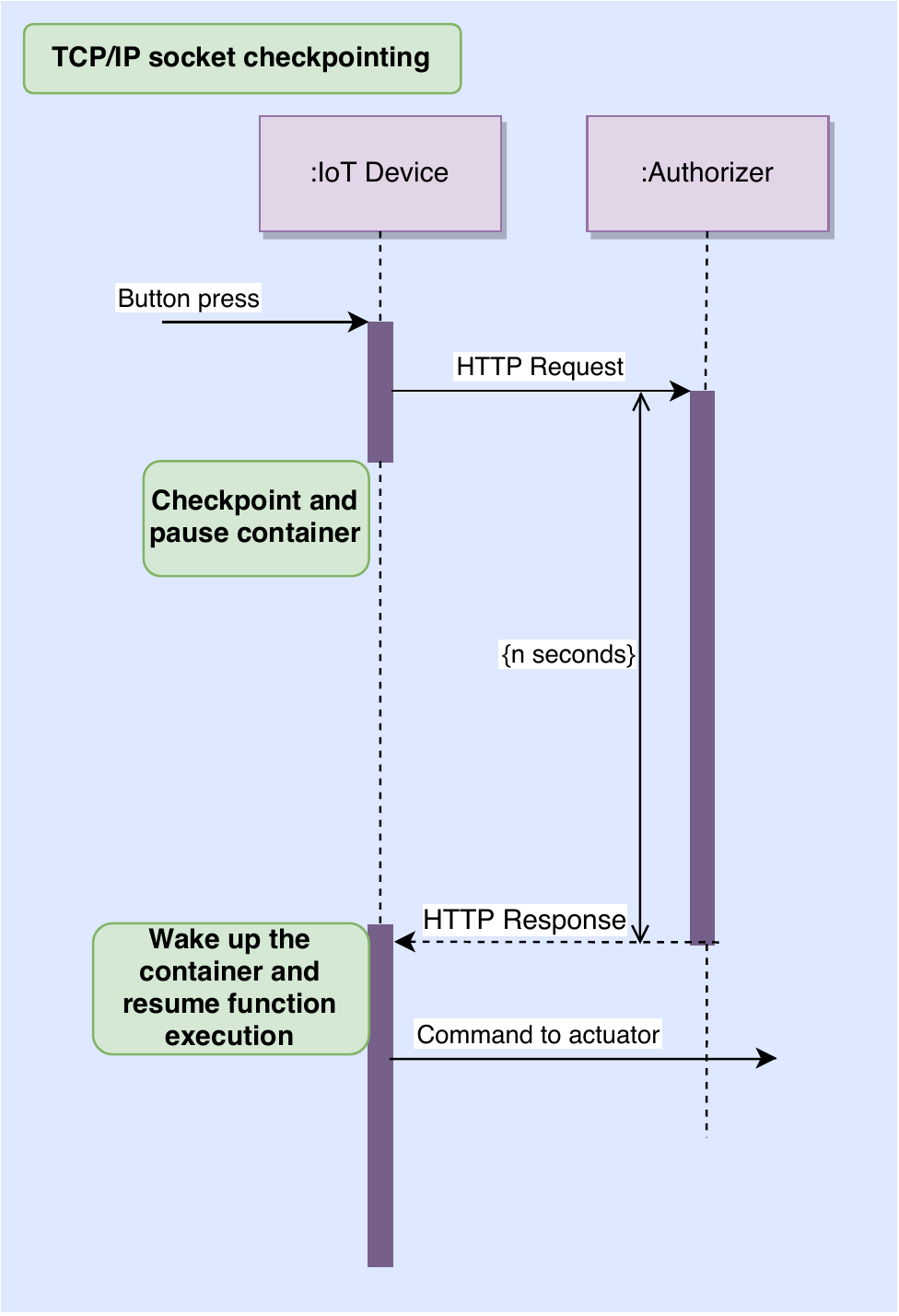}
    \caption{Checkpointing and resuming a TCP/IP socket in an HTTP request}
    \label{fig:tcpip}
\end{figure}

Fig. \ref{fig:tcpip} illustrates the stages in this test. First an IoT device receives a signal from a button or some other sensor device. The IoT device then sends a request to an authorizing entity for permission to use an actuator device. In this case, we assume that processing this request might take a long time, and therefore, we can put the container to sleep. It is essential to preserve the socket, which has an active connection to the server, so that we are able to receive the response message from the authorizing entity later. Once the authorization message has been sent and received by the IoT device, the IoT device wakes up the container and resumes the function execution. For practical purposes, we manually acknowledge the requests from the IoT devices, so the processing may take an arbitrary amount of time.

Checkpointing the TCP/IP connection in this example works without any modifications to CRIU or Docker. Although this scenario is somewhat simplified, it might have some powerful uses in saving system resources in a constrained IoT device. In some applications, it might make more sense to use a publish-subscribe message passing protocol, such as MQTT, due to its scalability and asynchronous mode of operation. This approach would also be compatible with MQTT, as MQTT also uses the TCP/IP protocol. The IoT device can implement long-running functions as MQTT subscriber modules. The subscriber could be put to sleep after registering to the broker and woken up by incoming events.

\subsubsection{Waking up the Function}

Linux Iptables can be used to trigger a wake-up call for a paused long-running function. We tested this approach with a Python script that listens for a packet arrival on a connection from a specific device and source port to a specific destination port on our IoT device. Upon the arrival of a matching packet, the script creates a new container and resumes the checkpointed function into it. The function receives the TCP/IP data and continues execution. We have been able to create the scenario of Fig. \ref{fig:tcpip} using this method.

\subsection{Migrating Container to Another IoT Device}

The goal of this experiment was to find out if it might be possible to checkpoint a function on one IoT device and resume it on another IoT device. We checkpoint the container state, send the checkpoint data to the target device, create a new container on the target device and restore it from the checkpoint. We have set up two Raspberry Pi devices with identical software packages and operating systems as mentioned in Section \ref{sec:setting-up-software}. However, the Raspberry Pis differ slightly in their hardware as the first device is version 3 Model B and the second device is version 2 Model B. 

%
%

We created a simple function in JavaScript/NodeJS containing a counter that was increased periodically. We were able to checkpoint the NodeJS function and container, migrate it to the second Raspberry Pi and verify that the counter continued where it left off. This result shows that it is possible to continue the function execution on another IoT device, provided that the target device has matching hardware and software architecture. Certainly more experience is needed from different kinds of functions as the script used in this test has very simple behaviour. The different hardware versions of the two devices did not cause any problems in this test.

The results obtained from this experiment hint at possible usage scenarios involving load balancing and fault-tolerant IoT devices as introduced in Section \ref{loadbalance} and Section \ref{fault}, respectively. Another interesting insight is that it would be possible to build systems that resemble mobile agent functionality, but using containers instead of agent frameworks that are often limited to a specific programming language or set of libraries. Mobile agents, although being an old concept, have been recently used in Wireless Sensor Networks and IoT with the idea that they can enable code mobility with very small memory footprint. However, mobile agents utilizing checkpointing would take another approach and improve programmability at the cost of memory footprint. Hence they would be targeting more computationally capable devices -- like the Raspberry Pis used in this example.

\subsection{Checkpointing Performance}

In this section, we present some early results from the implementation showing the feasibility of our approach.

\subsubsection{Image Sizes}

The unoptimized container size in the TCP/IP scenario was 257 MB including the NodeJS dependencies. The resulting checkpoint files were 15 MB in total. The checkpoint file size is sufficiently small for container live migration assuming that the target device has already loaded the container base image. This is an encouraging result for our approach of running FaaS-like functionality in the IoT devices, but more investigation is still needed.

\subsubsection{Container Latency Measurements}

The latency measurements were collected using the 'time' utility that comes with many GNU/Linux distributions or as a built-in keyword in the Bash shell. For our experiment, we used the Bash version. The results are presented in Table \ref{tab:freq}.

\begin{table}[h!]
  \caption{Checkpointing performance}
  \label{tab:freq}
  \begin{tabular}{lcl}
    \toprule
    Task&Time (s)\\
    \midrule
    Starting a container & 1.463\\
    Pause & 0.857\\
    Unpause & 0.850\\
    Create a checkpoint and save it to disk & 1.716\\
    Create a new container & 0.438\\
    Start container from the checkpoint & 1.763\\
  \bottomrule
\end{tabular}
\end{table}

As can be seen from the results, starting a container from the checkpoint takes about the same time as starting a fresh container. Also, creating a checkpoint takes around the same amount of time. These results are naturally different when the container is larger and more complex, but the time delay of starting a checkpointed container vs starting a new container should both grow linearly when increasing the container size.

Pausing and unpausing is a bit quicker way of saving the container state and continuing function execution, but we cannot use it for live migration or backing up the container state to disk. However, it could be usable in situations when we know that the function needs to run only on a specific device and nothing needs to be stored in the disk. 

\subsection{Memory Usage}

\begin{figure}[h!]
    \centering
    \includegraphics[height=0.6\linewidth, width=0.9\linewidth]{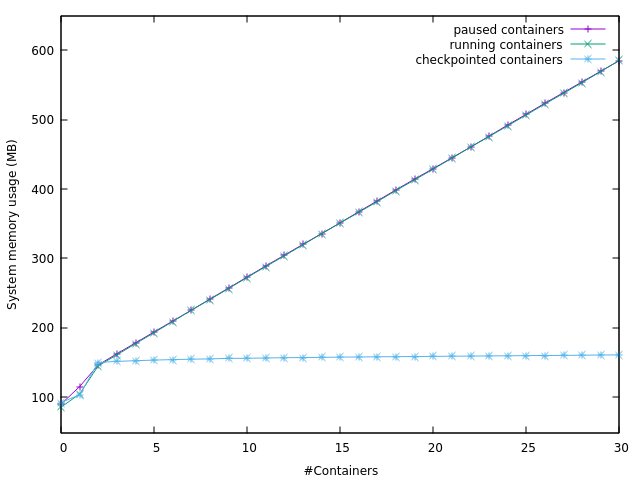}
    \caption{Memory footprint of running, paused and checkpointed containers}
    \label{fig:memory}
\end{figure}

Fig. \ref{fig:memory} shows the impact of active, paused and checkpointed Docker containers on the system memory. The linear lines of paused and running containers agree with the results obtained in \cite{hendrickson2016serverless}. However, in this scenario we are more limited with the available memory, as we run the tests on IoT devices. The checkpointed containers do not cause notable increase in system memory consumption, which was expected as the checkpoints are stored in the disk.

\section{Conclusion}
\label{conclusion}
In this paper, we have explored the function as a service concept applied to the IoT edge devices. We have utilized the checkpointing mechanism to suspend long-running blocking functions in order to save resources in constrained devices. We also successfully demonstrated live migration of containers using the combination of Docker and CRIU. The evaluation shows encouraging early results for the checkpointing functionality in IoT devices. These building blocks can be used to build fault-tolerant IoT systems, implement resource offloading as well as improve efficiency and availability in the IoT edge.


\begin{acks}
    The research work is conducted in the Wireless Innovation between Finland and the US (WiFiUS) Massive IoT project, and is supported by the Academy of Finland and the National Science Foundation (Grant No. 1702952). 


\end{acks}

%% file: sample-sigconf.bbl

\begin{thebibliography}{22}


\ifx \showCODEN    \undefined \def \showCODEN     #1{\unskip}     \fi
\ifx \showDOI      \undefined \def \showDOI       #1{#1}\fi
\ifx \showISBNx    \undefined \def \showISBNx     #1{\unskip}     \fi
\ifx \showISBNxiii \undefined \def \showISBNxiii  #1{\unskip}     \fi
\ifx \showISSN     \undefined \def \showISSN      #1{\unskip}     \fi
\ifx \showLCCN     \undefined \def \showLCCN      #1{\unskip}     \fi
\ifx \shownote     \undefined \def \shownote      #1{#1}          \fi
\ifx \showarticletitle \undefined \def \showarticletitle #1{#1}   \fi
\ifx \showURL      \undefined \def \showURL       {\relax}        \fi
\providecommand\bibfield[2]{#2}
\providecommand\bibinfo[2]{#2}
\providecommand\natexlab[1]{#1}
\providecommand\showeprint[2][]{arXiv:#2}

\bibitem[\protect\citeauthoryear{??}{cgr}{[n.d.]}]%
        {cgroupfreezer}
 \bibinfo{year}{[n.d.]}\natexlab{}.
\newblock \bibinfo{title}{{Cgroup Freezer}}.
\newblock
  \bibinfo{howpublished}{\url{https://www.kernel.org/doc/Documentation/cgroup-v1/freezer-subsystem.txt}}.
\newblock
\newblock
\shownote{Accessed: November 2018.}


\bibitem[\protect\citeauthoryear{A{\"{i}}ssaoui, Cooperman, Monteil, and
  Tazi}{A{\"{i}}ssaoui et~al\mbox{.}}{2016}]%
        {Aissaoui2016}
\bibfield{author}{\bibinfo{person}{Francois A{\"{i}}ssaoui},
  \bibinfo{person}{Gene Cooperman}, \bibinfo{person}{Thierry Monteil}, {and}
  \bibinfo{person}{Sa{\"{i}}d Tazi}.} \bibinfo{year}{2016}\natexlab{}.
\newblock \showarticletitle{{Smart Scene Management for IoT-based Constrained
  Devices Using Checkpointing}}.
\newblock \bibinfo{journal}{\emph{Proceedings - 2016 IEEE 15th International
  Symposium on Network Computing and Applications, NCA 2016}}
  (\bibinfo{year}{2016}), \bibinfo{pages}{170--174}.
\newblock
\showISBNx{9781509032167}
\urldef\tempurl%
\url{https://doi.org/10.1109/NCA.2016.7778613}
\showDOI{\tempurl}


\bibitem[\protect\citeauthoryear{Amazon}{Amazon}{[n.d.]a}]%
        {amazonlambda}
\bibfield{author}{\bibinfo{person}{Amazon}.}
  \bibinfo{year}{[n.d.]}\natexlab{a}.
\newblock \bibinfo{title}{{Amazon Lambda}}.
\newblock
  \bibinfo{howpublished}{\url{https://aws.amazon.com/lambda/resources/}}.
\newblock
\newblock
\shownote{Accessed: November 2018.}


\bibitem[\protect\citeauthoryear{Amazon}{Amazon}{[n.d.]b}]%
        {amazonlambdaedge}
\bibfield{author}{\bibinfo{person}{Amazon}.}
  \bibinfo{year}{[n.d.]}\natexlab{b}.
\newblock \bibinfo{title}{{Amazon Lambda$@$Edge}}.
\newblock \bibinfo{howpublished}{\url{https://aws.amazon.com/lambda/edge/}}.
\newblock
\newblock
\shownote{Accessed: January 2019.}


\bibitem[\protect\citeauthoryear{Apache}{Apache}{[n.d.]}]%
        {openwhisk}
\bibfield{author}{\bibinfo{person}{Apache}.} \bibinfo{year}{[n.d.]}\natexlab{}.
\newblock \bibinfo{title}{{Apache OpenWhisk}}.
\newblock \bibinfo{howpublished}{\url{https://openwhisk.apache.org/}}.
\newblock
\newblock
\shownote{Accessed: February 2019.}


\bibitem[\protect\citeauthoryear{Bellavista and Zanni}{Bellavista and
  Zanni}{2017}]%
        {bellavista:rpifeasibility}
\bibfield{author}{\bibinfo{person}{Paolo Bellavista} {and}
  \bibinfo{person}{Alessandro Zanni}.} \bibinfo{year}{2017}\natexlab{}.
\newblock \showarticletitle{Feasibility of Fog Computing Deployment Based on
  Docker Containerization over RaspberryPi}. In
  \bibinfo{booktitle}{\emph{Proceedings of the 18th International Conference on
  Distributed Computing and Networking}} (Hyderabad, India)
  \emph{(\bibinfo{series}{ICDCN '17})}. \bibinfo{publisher}{ACM},
  \bibinfo{address}{New York, NY, USA}, Article \bibinfo{articleno}{16},
  \bibinfo{numpages}{10}~pages.
\newblock
\showISBNx{978-1-4503-4839-3}
\urldef\tempurl%
\url{https://doi.org/10.1145/3007748.3007777}
\showDOI{\tempurl}


\bibitem[\protect\citeauthoryear{Chen}{Chen}{2015}]%
        {Chen2015}
\bibfield{author}{\bibinfo{person}{Yang Chen}.}
  \bibinfo{year}{2015}\natexlab{}.
\newblock \showarticletitle{{Checkpoint and Restore of Micro-service in Docker
  Containers}}.
\newblock \bibinfo{journal}{\emph{Proceedings of the 3rd International
  Conference on Mechatronics and Industrial Informatics}}
  \bibinfo{number}{Icmii} (\bibinfo{year}{2015}), \bibinfo{pages}{915--918}.
\newblock
\showISBNx{978-94-6252-131-5}
\urldef\tempurl%
\url{https://doi.org/10.2991/icmii-15.2015.160}
\showDOI{\tempurl}


\bibitem[\protect\citeauthoryear{CRIU}{CRIU}{[n.d.]a}]%
        {criu}
\bibfield{author}{\bibinfo{person}{CRIU}.} \bibinfo{year}{[n.d.]}\natexlab{a}.
\newblock \bibinfo{title}{{CRIU Checkpoint/Restore Functionality}}.
\newblock \bibinfo{howpublished}{\url{https://criu.org/Main_Page}}.
\newblock
\newblock
\shownote{Accessed: November 2018.}


\bibitem[\protect\citeauthoryear{CRIU}{CRIU}{[n.d.]b}]%
        {criukernel}
\bibfield{author}{\bibinfo{person}{CRIU}.} \bibinfo{year}{[n.d.]}\natexlab{b}.
\newblock \bibinfo{title}{{Kernel Options}}.
\newblock \bibinfo{howpublished}{\url{https://criu.org/Linux_kernel}}.
\newblock
\newblock
\shownote{Accessed: November 2018.}


\bibitem[\protect\citeauthoryear{Docker}{Docker}{[n.d.]}]%
        {dockerpause}
\bibfield{author}{\bibinfo{person}{Docker}.} \bibinfo{year}{[n.d.]}\natexlab{}.
\newblock \bibinfo{title}{{Docker Pause}}.
\newblock
  \bibinfo{howpublished}{\url{https://docs.docker.com/engine/reference/commandline/pause/\#usage}}.
\newblock
\newblock
\shownote{Accessed: November 2018.}


\bibitem[\protect\citeauthoryear{Google}{Google}{[n.d.]}]%
        {googlecloud}
\bibfield{author}{\bibinfo{person}{Google}.} \bibinfo{year}{[n.d.]}\natexlab{}.
\newblock \bibinfo{title}{{Google Cloud Functions}}.
\newblock
  \bibinfo{howpublished}{\url{https://cloud.google.com/functions/docs/}}.
\newblock
\newblock
\shownote{Accessed: November 2018.}


\bibitem[\protect\citeauthoryear{Hellerstein, Faleiro, Gonzalez,
  Schleier-Smith, Sreekanti, Tumanov, and Wu}{Hellerstein
  et~al\mbox{.}}{2018}]%
        {hellerstein2018serverless}
\bibfield{author}{\bibinfo{person}{Joseph~M Hellerstein}, \bibinfo{person}{Jose
  Faleiro}, \bibinfo{person}{Joseph~E Gonzalez}, \bibinfo{person}{Johann
  Schleier-Smith}, \bibinfo{person}{Vikram Sreekanti}, \bibinfo{person}{Alexey
  Tumanov}, {and} \bibinfo{person}{Chenggang Wu}.}
  \bibinfo{year}{2018}\natexlab{}.
\newblock \showarticletitle{Serverless Computing: One Step Forward, Two Steps
  Back}.
\newblock \bibinfo{journal}{\emph{arXiv preprint arXiv:1812.03651}}
  (\bibinfo{year}{2018}).
\newblock


\bibitem[\protect\citeauthoryear{Hendrickson, Sturdevant, Harter,
  Venkataramani, Arpaci-Dusseau, and Arpaci-Dusseau}{Hendrickson
  et~al\mbox{.}}{2016}]%
        {hendrickson2016serverless}
\bibfield{author}{\bibinfo{person}{Scott Hendrickson}, \bibinfo{person}{Stephen
  Sturdevant}, \bibinfo{person}{Tyler Harter}, \bibinfo{person}{Venkateshwaran
  Venkataramani}, \bibinfo{person}{Andrea~C Arpaci-Dusseau}, {and}
  \bibinfo{person}{Remzi~H Arpaci-Dusseau}.} \bibinfo{year}{2016}\natexlab{}.
\newblock \showarticletitle{Serverless computation with openlambda}. In
  \bibinfo{booktitle}{\emph{8th $\{$USENIX$\}$ Workshop on Hot Topics in Cloud
  Computing (HotCloud 16)}}.
\newblock


\bibitem[\protect\citeauthoryear{Kakakhel, Mukkala, Westerlund, and
  Plosila}{Kakakhel et~al\mbox{.}}{2018}]%
        {Kakakhel2018}
\bibfield{author}{\bibinfo{person}{Syed Rameez~Ullah Kakakhel},
  \bibinfo{person}{Lauri Mukkala}, \bibinfo{person}{Tomi Westerlund}, {and}
  \bibinfo{person}{Juha Plosila}.} \bibinfo{year}{2018}\natexlab{}.
\newblock \showarticletitle{{Virtualization at the Network Edge: A Technology
  Perspective}}.
\newblock \bibinfo{journal}{\emph{2018 3rd International Conference on Fog and
  Mobile Edge Computing, FMEC 2018}} (\bibinfo{year}{2018}),
  \bibinfo{pages}{87--92}.
\newblock
\showISBNx{9781538658963}
\urldef\tempurl%
\url{https://doi.org/10.1109/FMEC.2018.8364049}
\showDOI{\tempurl}


\bibitem[\protect\citeauthoryear{Mehta, Baddour, Svensson, Gustafsson, and
  Elmroth}{Mehta et~al\mbox{.}}{2017}]%
        {Mehta2017}
\bibfield{author}{\bibinfo{person}{Amardeep Mehta}, \bibinfo{person}{Rami
  Baddour}, \bibinfo{person}{Fredrik Svensson}, \bibinfo{person}{Harald
  Gustafsson}, {and} \bibinfo{person}{Erik Elmroth}.}
  \bibinfo{year}{2017}\natexlab{}.
\newblock \showarticletitle{{Calvin Constrained -- A Framework for IoT
  Applications in Heterogeneous Environments}}. In
  \bibinfo{booktitle}{\emph{2017 IEEE 37th International Conference on
  Distributed Computing Systems (ICDCS)}}. \bibinfo{publisher}{IEEE},
  \bibinfo{pages}{1063--1073}.
\newblock
\showISBNx{978-1-5386-1792-2}
\urldef\tempurl%
\url{https://doi.org/10.1109/ICDCS.2017.181}
\showDOI{\tempurl}


\bibitem[\protect\citeauthoryear{Microsoft}{Microsoft}{[n.d.]a}]%
        {microsoftazure}
\bibfield{author}{\bibinfo{person}{Microsoft}.}
  \bibinfo{year}{[n.d.]}\natexlab{a}.
\newblock \bibinfo{title}{{Azure Functions}}.
\newblock
  \bibinfo{howpublished}{\url{https://docs.microsoft.com/en-us/azure/azure-functions}}.
\newblock
\newblock
\shownote{Accessed: November 2018.}


\bibitem[\protect\citeauthoryear{Microsoft}{Microsoft}{[n.d.]b}]%
        {azureedge}
\bibfield{author}{\bibinfo{person}{Microsoft}.}
  \bibinfo{year}{[n.d.]}\natexlab{b}.
\newblock \bibinfo{title}{{Azure IoT Edge}}.
\newblock
  \bibinfo{howpublished}{\url{https://docs.microsoft.com/en-us/azure/iot-edge/}}.
\newblock
\newblock
\shownote{Accessed: January 2019.}


\bibitem[\protect\citeauthoryear{Microsoft}{Microsoft}{[n.d.]c}]%
        {durablefunctionsfw}
\bibfield{author}{\bibinfo{person}{Microsoft}.}
  \bibinfo{year}{[n.d.]}\natexlab{c}.
\newblock \bibinfo{title}{{Durable Functions Framework}}.
\newblock
  \bibinfo{howpublished}{\url{https://docs.microsoft.com/en-us/azure/azure-functions/durable-functions-overview\#the-technology}}.
\newblock
\newblock
\shownote{Accessed: November 2018.}


\bibitem[\protect\citeauthoryear{Mirhoseini, Rouhani, Songhori, and
  Koushanfar}{Mirhoseini et~al\mbox{.}}{2016}]%
        {Mirhoseini2016}
\bibfield{author}{\bibinfo{person}{Azalia Mirhoseini},
  \bibinfo{person}{Bita~Darvish Rouhani}, \bibinfo{person}{Ebrahim Songhori},
  {and} \bibinfo{person}{Farinaz Koushanfar}.} \bibinfo{year}{2016}\natexlab{}.
\newblock \showarticletitle{{Chime: Checkpointing Long Computations on
  Intermittently Energized IoT Devices}}.
\newblock \bibinfo{journal}{\emph{IEEE Transactions on Multi-Scale Computing
  Systems}} \bibinfo{volume}{2}, \bibinfo{number}{4} (\bibinfo{year}{2016}),
  \bibinfo{pages}{277--290}.
\newblock
\showISBNx{VO - 2}
\showISSN{23327766}
\urldef\tempurl%
\url{https://doi.org/10.1109/TMSCS.2016.2550442}
\showDOI{\tempurl}


\bibitem[\protect\citeauthoryear{Morabito}{Morabito}{2017}]%
        {Morabito2017}
\bibfield{author}{\bibinfo{person}{Roberto Morabito}.}
  \bibinfo{year}{2017}\natexlab{}.
\newblock \showarticletitle{{Virtualization on Internet of Things Edge Devices
  With Container Technologies: A Performance Evaluation}}.
\newblock  (\bibinfo{year}{2017}), \bibinfo{pages}{8835--8850}.
\newblock


\bibitem[\protect\citeauthoryear{Qiu, Lung, Ajila, and Srivastava}{Qiu
  et~al\mbox{.}}{2017}]%
        {Qiu2017}
\bibfield{author}{\bibinfo{person}{Yuqing Qiu}, \bibinfo{person}{Chung~Horng
  Lung}, \bibinfo{person}{Samuel Ajila}, {and} \bibinfo{person}{Pradeep
  Srivastava}.} \bibinfo{year}{2017}\natexlab{}.
\newblock \showarticletitle{{LXC Container Migration in Cloudlets under
  Multipath TCP}}.
\newblock \bibinfo{journal}{\emph{Proceedings - International Computer Software
  and Applications Conference}}  \bibinfo{volume}{2} (\bibinfo{year}{2017}),
  \bibinfo{pages}{31--36}.
\newblock
\showISBNx{9781538603673}
\showISSN{07303157}
\urldef\tempurl%
\url{https://doi.org/10.1109/COMPSAC.2017.163}
\showDOI{\tempurl}


\bibitem[\protect\citeauthoryear{Wang, Li, Zhang, Ristenpart, and Swift}{Wang
  et~al\mbox{.}}{2018}]%
        {wang2018peeking}
\bibfield{author}{\bibinfo{person}{Liang Wang}, \bibinfo{person}{Mengyuan Li},
  \bibinfo{person}{Yinqian Zhang}, \bibinfo{person}{Thomas Ristenpart}, {and}
  \bibinfo{person}{Michael Swift}.} \bibinfo{year}{2018}\natexlab{}.
\newblock \showarticletitle{Peeking Behind the Curtains of Serverless
  Platforms}. In \bibinfo{booktitle}{\emph{2018 {USENIX} Annual Technical
  Conference ({USENIX} {ATC} 18)}}. \bibinfo{publisher}{{USENIX} Association},
  \bibinfo{address}{Boston, MA}, \bibinfo{pages}{133--146}.
\newblock
\showISBNx{978-1-931971-44-7}
\urldef\tempurl%
\url{https://www.usenix.org/conference/atc18/presentation/wang-liang}
\showURL{%
\tempurl}


\end{thebibliography}
